\documentclass[12pt,journal]{IEEEtran}
\usepackage{graphicx}
\ifCLASSOPTIONcompsoc
\else
\fi
\ifCLASSINFOpdf
\else
\fi

\begin{document}

\title{Portable Tor Router:\\ Easily Enabling Web Privacy for Consumers}
%\title{The Tor Router: A Simple Solution to Greater On-line Privacy}

\author{\IEEEauthorblockN{
        Adrian~Barberis,
        Danny~Radosevich,
        Wyatt~Emery,
        and Mike Borowczak,\textit{ Member, IEEE}}\\
\IEEEauthorblockA{Computer Science Department, University of Wyoming\\
dradose1, abarber3, wemery1, mborowcz @uwyo.edu}
}

\IEEEcompsoctitleabstractindextext{%
\begin{abstract}
%\boldmath
On-line privacy is of major public concern. Unfortunately, for the average consumer, there is no simple mechanism to browse the Internet privately on multiple devices. Most available Internet privacy mechanisms are either expensive, not readily available, untrusted, or simply provide trivial information masking. We propose that the simplest, most effective and inexpensive way of gaining privacy, without sacrificing unnecessary amounts of functionality and speed, is to mask the user's IP address while also encrypting all data. We hypothesized that the Tor protocol is aptly suited to address these needs.  With this in mind we implemented a Tor router using a single board computer and the open-source Tor protocol code. We found that our proposed solution was able to meet five of our six goals soon after its implementation: cost effectiveness, immediacy of privacy, simplicity of use, ease of execution, and unimpaired functionality. Our final criterion of speed was sacrificed for greater privacy but it did not fall so low as to impair day-to-day functionality.  With a total cost of roughly \$100.00 USD and a speed cap of around 2 Megabits per second we were able to meet our goal of an affordable, convenient, and usable solution to increased on-line privacy for the average consumer.
\end{abstract}
% not want either math or citations to appear in the abstract.

}

\maketitle

% is not selected <OR> if conference mode is selected - because compsoc
% conference papers position the abstract like regular (non-compsoc)
\IEEEdisplaynotcompsoctitleabstractindextext

% \fi
\IEEEpeerreviewmaketitle

\section{Introduction}
% Computer Society journal papers do something a tad strange with the very
% first section heading (almost always called "Introduction"). They place it
% ABOVE the main text! IEEEtran.cls currently does not do this for you.
% However, You can achieve this effect by making LaTeX jump through some
% hoops via something like:
%
%\ifCLASSOPTIONcompsoc
%  \noindent\raisebox{2\baselineskip}[0pt][0pt]%
%  {\parbox{\columnwidth}{\section{Introduction}\label{sec:introduction}%
%  \global\everypar=\everypar}}%
%  \vspace{-1\baselineskip}\vspace{-\parskip}\par
%\else
%  \section{Introduction}\label{sec:introduction}\par
%\fi
%
% Admittedly, this is a hack and may well be fragile, but seems to do the
% trick for me. Note the need to keep any \label that may be used right
% after \section in the above as the hack puts \section within a raised box.

\IEEEPARstart {}{} As more and more data is generated, tracked, collected, bought, and sold, an individual's on-line privacy becomes more and more important. Although the data payload of any given transmission will be (or should be) encrypted, the header file which is used to route the data to its destination is not.  This allows an intermediary to scan and track a users data packet from source to destination easily and can compromise the users privacy by revealing information such as the IP address of the user, destination IP address (which can clue the intermediary in on what the user is doing while on-line), and time of use (which gives the intermediary an idea of when the user gets on-line on a given day). We decided to see how feasible it would be to apply an extra layer of privacy (without sacrificing so much speed as to impede normal on-line operations) to daily Internet activities; revealing the Tor protocol as a possible solution. We wish to determine the practicality of a Tor router in an attempt to provide the average user a little more peace of mind. Tor routing provides an extra layer of security while browsing the Internet by masking a packet's path through Onion Routing, a routing protocol encapsulating data in layers of encryption. This protects the user from attackers as well as taking away an ISP's ability to gather and sell a user's browsing data. The purpose of our project was to build a Tor router, conduct speed test comparisons, and research possible vulnerabilities which would lead to loss of privacy.  We specifically focused on a Tor router, as opposed to the Tor browser, due to its portability and versatility across multiple kinds if devices. In order to paint a broader picture of the functionality of such a device, we will also discuss how a Tor router may be used by the average user to increase privacy and why people should consider adopting this technology.  

\section{Background}

\subsection{Current Work Regarding Tor Networks}
	The Tor Project deals with increasing the level of privacy afforded to a user \cite{snader2008tune}\cite{reardon2009improving} and designing new versions of the Tor protocol that the network uses \cite{rochet2017waterfilling}\cite{kale2016improving}. The hope is to increase overall throughput and speed as well as eliminate DNS leaks and other potential privacy leaks through the use of advanced data mining\cite{oda2017user}. Other work includes researching how the Tor network is being used, how it impacts the user's experience, and how it impacts the Internet. A 2016 study by Andrea Forte, Nazanin Andalibi and Rachel Greenstadt \cite{Konr10cm} sought to understand how websites that block Tor users affect the on-line experiences of those that wish to maintain their privacy but also contribute to the on-line knowledge pool, specifically Wikipedia. Other more technical avenues of research include the work done by the maintainers of \textit{Tonga}, the Tor network's \textit{Bridge Authority}, which endeavors to reduce limitations in current implementations of the Tor code. Although of great importance, such research is unconcerned with Tor's viability as a consumer related Internet privacy system.

\subsection{The Tor Protocol}
\begin{figure}[htb]
  \includegraphics[width=0.5\textwidth]{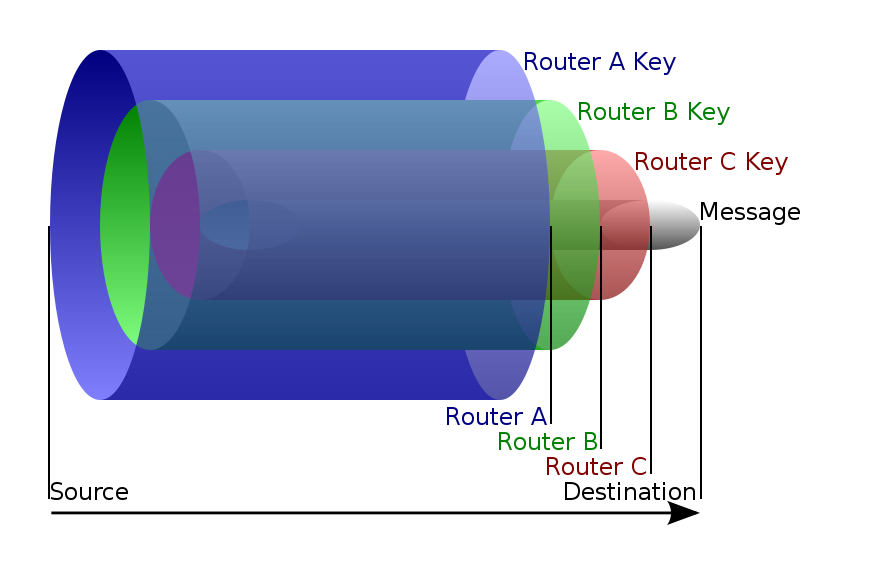}
  \caption{A Depiction of Onion Routing\cite{Konr06cm}.}
  \label{fig:Onion}
\end{figure}

Tor is built off of a protocol called \textit{Onion Routing} which was created by DARPA and patented by the US Navy in 1998 \cite{OnionRouting}. The name is derived from the use of multiple encryption layers applied to a packet when sending from a source to a destination node. (Figure \ref{fig:Onion} depicts a packet constructed by a user who is using onion routing.) The prime directive of onion routing and the Tor network is to hide as much information about a user and their data as possible from each interconnected communication node. Connection set up begins when the sender's Tor client (originator) asks a directory server (directory node) for a list of Tor nodes. It then proceeds to pick a path (chain) through the Tor nodes to the receiver's node (exit node).  The originator then asks the directory node for a public key, which is generated using asymmetric key cryptography.  Using this key, the user's Tor client can begin a encrypted connection to the first node (entry node) which will initiate an encrypted connection to the second node, and so on down the chain (a shared session key ensures that the connection is authentic).  The message to be sent from a node $X$ is then encrypted by the originator, such that only node $X+1$ can decrypt it (e.g. if the path length is four  nodes, then the innermost encryption layer will be decipherable only by node four, while the next layer up is decipherable only by node three, and so on) \cite{Konr08cm}.  Additionally, even though each added node technically extends the originator's encrypted connection, each node along the chain does not know anything about the previous or next node except that they are the "previous" and "next" nodes. This form of "hopping" from node to node, with each node knowing nothing about the overall chain, ensures that the path of a packet cannot be discerned by an outside party\cite{Konr07cm}.

% If my exlpanation is too hard to follow we can fuse it with the one below
%When the user wishes to send a packet to a destination, it creates something similar to what is shown in the image. The message is the regular packet that would be sent without using onion routing.  From there the user will encrypt, using the proper keys, the message with C’s key, followed by B’s then A’s. Then the packet will be sent off to the entry node, or Router A. Router A will decrypt the first layer and send the packet on to B. B will do the same and send it on to C. Finally, C will decrypt the final layer and send the packet on to the final destination. This lack of trust between nodes helps ensure privacy throughout the network.\cite{Konr08cm}

Although, during transit, the packet is hidden, weaknesses remain at the entry and exit nodes \cite{alsabah2016performance}.  This is due to the fact that both the entry and exit nodes are aware of their location in the chain and are the only nodes which are connected to non-Tor networks. The chance of gaining any information from a compromised entry node is minimal due to the multiple layers of encryption that are applied to the packet. However, that does mean that an intermediary could capture other seemingly less sensitive information, such as the size of a packet being sent or the fact that a user is using the Tor protocol. This could give the intermediary a clue as to what is going on (consider oppressive governments that block all Tor network entry nodes). However, the information gathered is far less than what could normally be captured while not using the Tor protocol. Regardless, some usable data is still left out in the open. This has led to a hidden system of Tor entry nodes allowing users to connect to the Tor network without the attacker's foreknowledge. While the entry node does not pose a significant threat of privacy leaks, the exit node is an altogether different story. At the exit node, all encryption afforded by onion routing ends and what comes out of the exit node is readable plain-text.  This is catastrophic as it not only lets anyone see the packet's contents, but it also makes known the packet's final destination, whether innocent or illicit.  However, the true source remains obscured. Luckily, this weakness can be minimized by using end-to-end encryption (such as using HTTPS vs HTTP) \cite{Konr07cm}.

\section{Using Tor for Consumer Privacy}
In this work, the primary objective is to explore a privacy solution for the general user or consumer that is both feasible and functional. In creating a small form-factor Wi-Fi router that routes all traffic through the Tor network, this work enables a portable and easy way to obscure user data while connecting devices to the Internet. The device is compact and can utilize either Ethernet or Wi-Fi to connect to a network. The router provides all services that a normal router would, including DNS, DHCP, and NAT. Although, there are a few drawbacks, the Tor router provides a compromise between efficiency and security. In order to validate the use of a Tor-enabled router, efficiency was measured by evaluating the impact on download speeds, security was measured by performing a DNS leak test. Implementation and test methodology are found in the following sections. 

\subsection{Implementation}

The set-top Tor router was implemented using a Raspberry Pi 3 running Raspbian OS. The assembled Raspberry Pi was then housed within a plastic casing for protection. The Raspian Stretch Lite OS, which does not include a desktop environment, was selected in order to maximize performance and speed of the Tor Router. Subsequently, the Raspberry Pi was configured to act as a normal router to perform all standard routing tasks. The Raspberry Pi's internal Wi-Fi card was then configured to connect to the local network.  Additionally, an externally connected Wi-Fi dongle was then configured an to broadcast an SSID, and act as an access point which was encrypted via WPA/WPA2. With the set-top router able to accept connections from hosts, a local DHCP server was enabled on the device to handle assigning IP addresses to connected hosts. Finally, DNS services were provided by using a Google Public DNS Server. At this point, the set-top Raspberry Pi was a fully functioning router with one important difference; whereas, most routers connect using only Ethernet this router was also able to connect via Wi-Fi. This makes it far more portable than a standard router. 

With a functional router in hand, the next main objective was to force all packets to be routed through the Tor network instead of directly to their destination. In order to achieve this, Tor software was installed on the Raspberry Pi and configured to run automatically on startup. All of the routing services occur on the Wi-Fi interface that accepts incoming connections from hosts. The outgoing interface is constantly connected to the Tor network. This creates a situation where all packets that exit the Pi , regardless of their original source, are sent straight to the Tor network, providing on-line privacy regardless of which application the user may be running.
\section{Results}

\subsection{Impact on Speed}
\begin{figure}[htb]
  \includegraphics[width=0.52\textwidth]
  {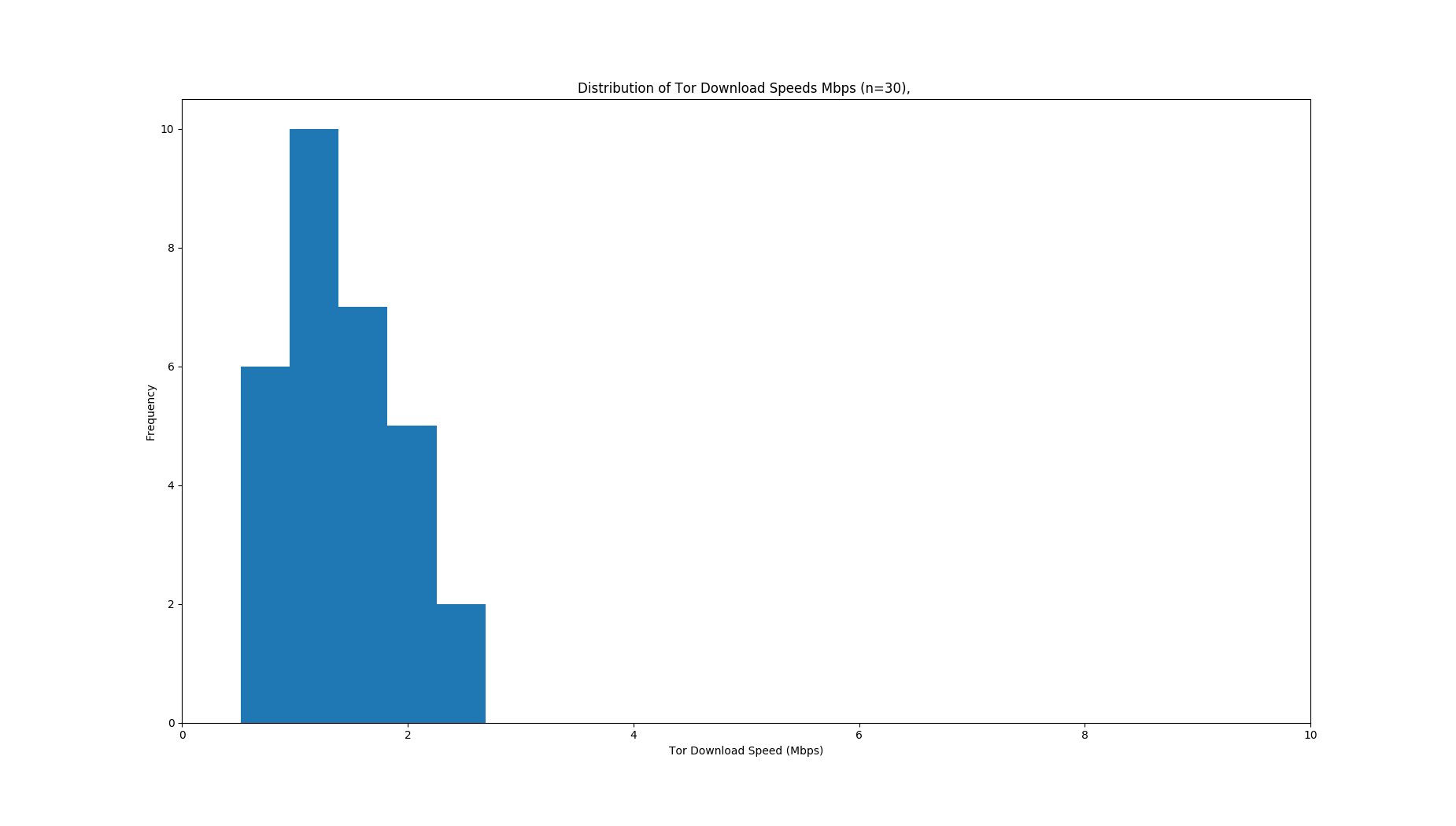}
  \caption{Internet download speed (Mb/s) on Wi-Fi \textbf{with} Tor router.}
  \label{fig:DOWNTOR}
\end{figure}

\begin{figure}[htb]
  \includegraphics[width=0.52\textwidth]
  {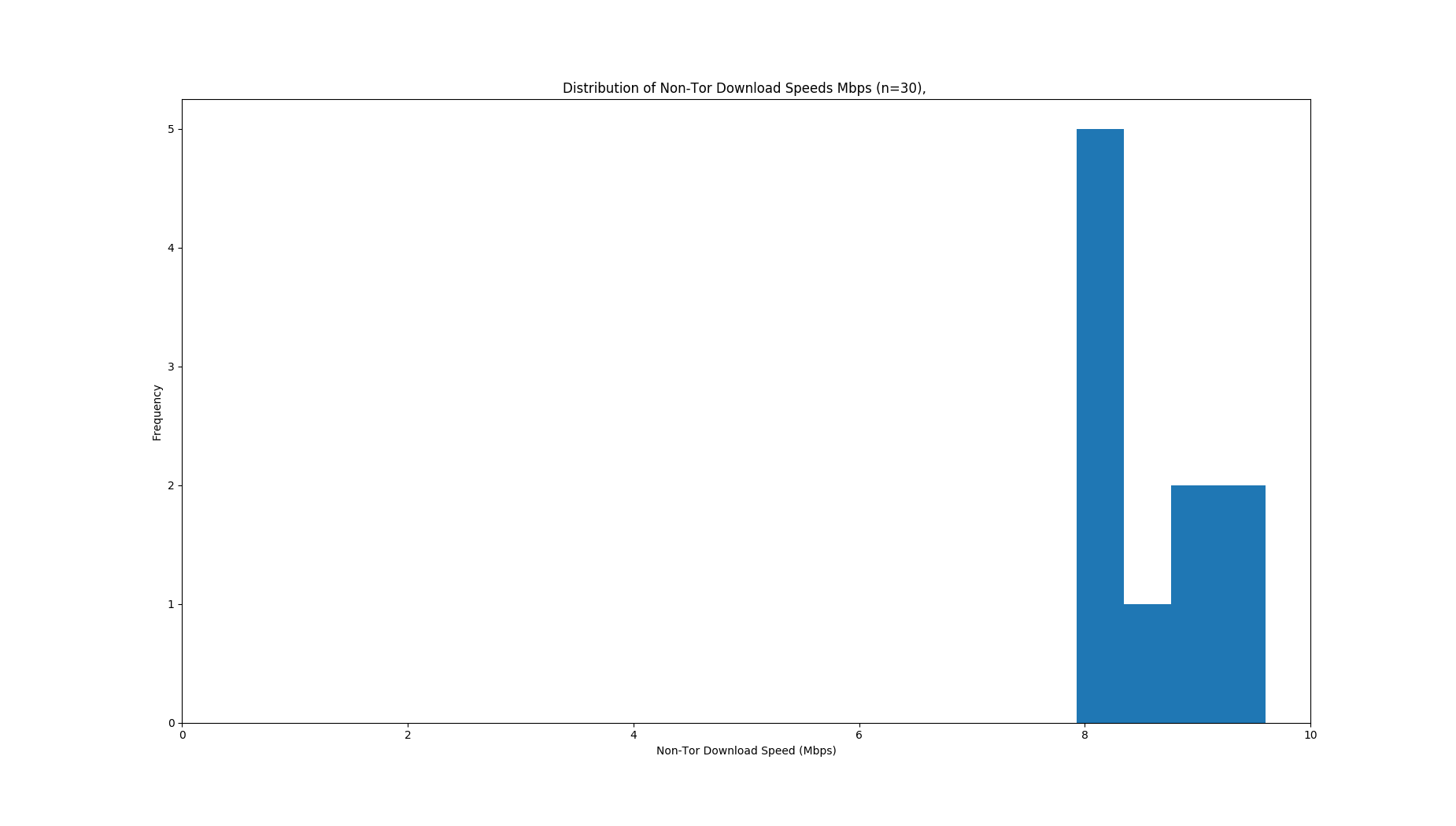}
  \caption{Internet download speed (Mb/s) on Wi-Fi \textbf{without} Tor router.}
  \label{fig:DOWNNOTOR}
\end{figure}

\begin{figure}[htb]
  \includegraphics[width=0.52\textwidth]
  {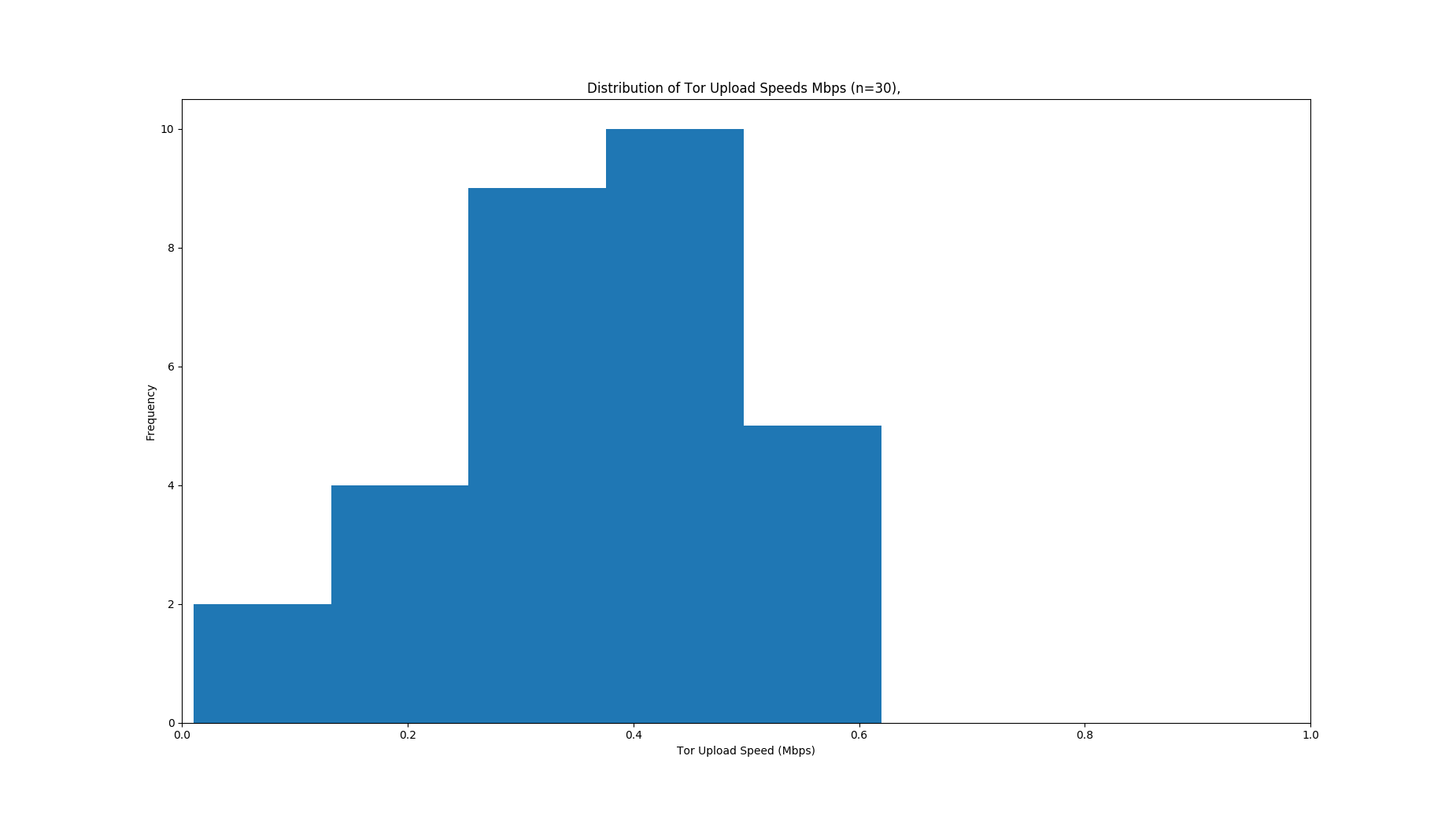}
  \caption{Internet upload speed (Mb/s) on Wi-Fi \textbf{with} Tor router.}
  \label{fig:UPTOR}
\end{figure}

\begin{figure}[htb]
  \includegraphics[width=0.52\textwidth]
  {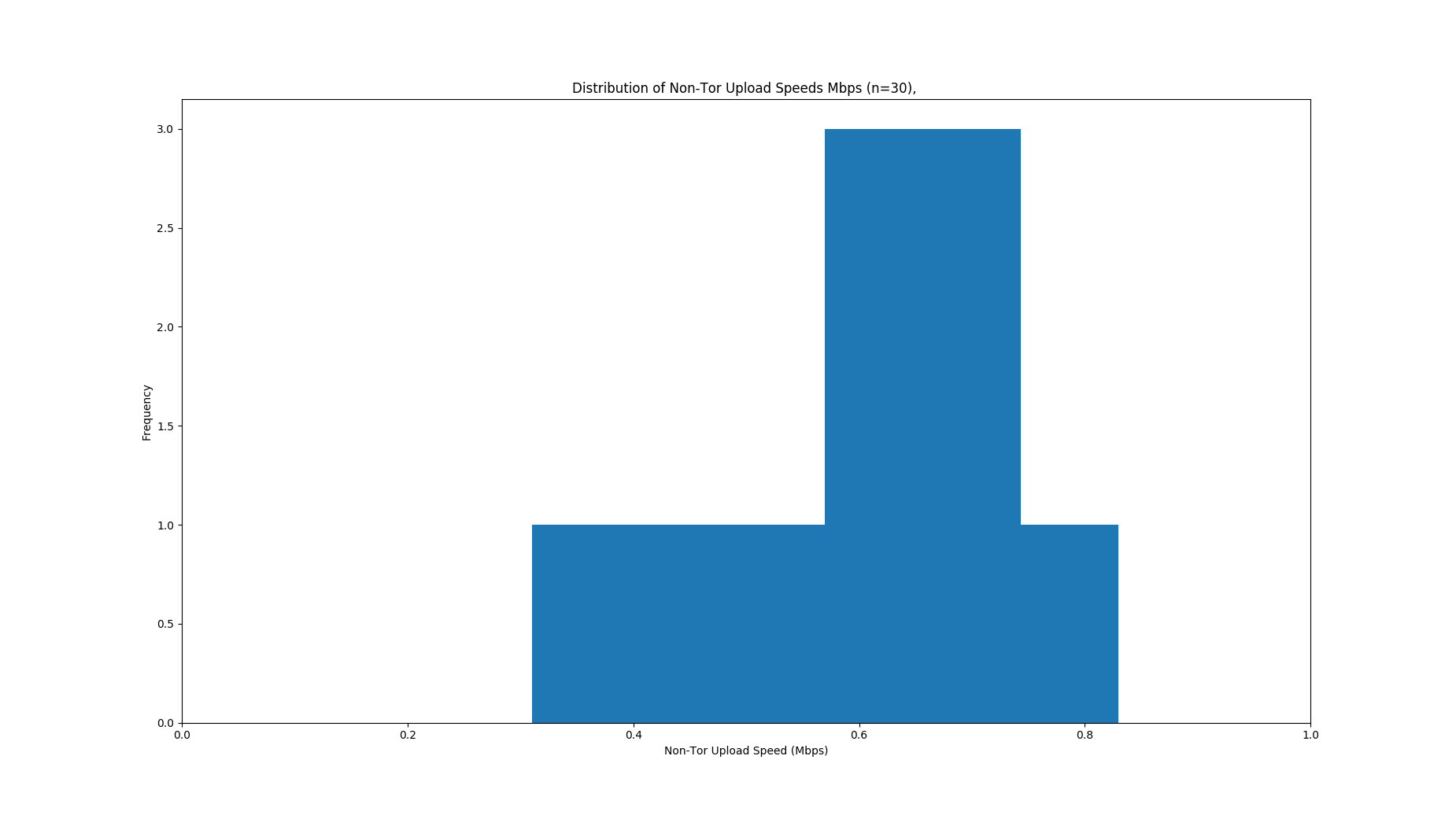}
  \caption{Internet upload speed (Mb/s) on Wi-Fi \textbf{without} Tor router.}
  \label{fig:UPNOTOR}
\end{figure}

In order to determine the impact on speed, multiple upload and download speed tests were performed using the Tor box router - with and without a Tor enabled connection. The results varied depending on the path through which packets were routed. The greater the distance, the greater impact on the speed.  In general, upload speed seemed to vary more than download speed which stayed between 0.5Mb/s and 2Mb/s with around 1Mb/s to 1.5Mb/s being the most commonly seen download speed. Figure \ref{fig:DOWNTOR} shows the distribution of download speeds while using Tor (avg. 1.7 MB/s) while Figure \ref{fig:DOWNNOTOR} shows the distribution of download speeds without use of Tor (avg. 8.3 MB/s). The same analysis was performed for upload speeds with results shown for Tor Upload and Non-Tor uploads in Figures \ref{fig:UPTOR} and \ref{fig:UPNOTOR} respectively.
While at first glance these values seem to contradict our axiom of maintaining usability, they are, in reality, perfectly functional and usable speeds.  In addition to the formal speed tests, while connected to the Tor network, we were able to stream video at resolutions up to 720p as well as reach and interact with various websites.  All of which are day-to-day activities an average user might engage in.  However, it was near impossible to download large files or stream full high definition video content.  

This variance in speed is expected due to the overhead of multiple encryptions and decryption as well as the often very roundabout way in which a packet is routed.  Additionally, the overall speed of the connection will invariably get throttled by the slowest node in the chain, a problem that can only be resolved by increasing throughput for all Tor servers. As is the case with many things, the efficiency is reduced in an effort to increase security.

\vspace{40 mm}

\subsection{Domain Name Server (DNS) Queries}
We also performed a DNS leak test in order to verify that the Tor router was in fact handling DNS queries properly. This is crucial to securing a user's browsing data.  When an originator does not use the Tor network's DNS services to complete DNS requests a DNS leak occurs. DNS leaks allow for an outside party to observe the websites a user has visited. When a user makes a DNS request to find the IP address of a website they wish to visit, they normally use their local DNS server, which then may contact a root DNS server, followed by a Top Level Domain server, followed by an authoritative server.  Each of the connections to these servers can be eavesdropped on in order gain more information a user's on-line habits.  By using the regular DNS servers for website queries, a user completely forgoes the Tor network's encryption to discover the IP address of a website they wish to visit. While this shows nothing about what parts of a website a user has visited, it does show where they have been or where they want to go. Such information can be used to profile a user.  The DNS leak test showed that our router was not leaking DNS requests. Instead, it used the Tor network's DNS services to complete DNS requests.

\section{Limitations}

Onion routing adds layers of encryption and security to a packet, allowing for anonymous packet transmission. The entry relay passes to the middle relay, which passes to the exit relay. Each one only knows how to communicate with its direct neighbors. Unencrypted text is sent out the exit relay toward the final destination and one must take further steps to secure their privacy beyond that point. Secret entry nodes exist to allow people to connect to the Tor network where governments and/or other entities have banned its use. Packets can be sniffed before reaching the entry relay but decrypting them is generally not a feasible task. Tor also places minimal trust in each node. Therefore, even if a node is compromised along the way, it is highly unlikely the attacker will be able to determine your real IP address. 

While the router discussed in this work solves some of the issues associated with using Tor, it also introduces other problems. Everyone connected to the router has not yet reached the Tor network and all internal network traffic is visible. Additionally, extra care must be taken to prevent compromise of the router, otherwise any chance of privacy is eliminated, unless one uses the Tor browser in conjunction with the router. Use of the Tor browser obscures endpoint identification and packet-sniffing by others connected to the router.

In addition to the limitations of the Tor Router, according to the Tor Project there are several common actions that may compromise an individual's privacy while on a Tor Network \cite{Konr09cm}. These actions are summarized into the following two broad categories: system behavior and browsing behavior. Examples of system-wide behavior that can impact anonymity include using other programs on your system and not using the Tor Browser. Each of these behaviors leaks out user information that is not necessarily related to browsing contents - but rather fingerprints of a user's system. Examples of browsing specific behaviors include downloading torrents, using browser plug-ins, and not using HTTPS. Each of these behaviors mitigates and potentially circumvents the privacy afforded to a user while using a Tor browser and router combination.

\section{Discussion and Conclusion}

When used properly, our Tor router provides a secure way for the average user to browse the Internet. While there are some drawbacks, we have shown that Tor does secure a user's browsing data well enough to shield the user from unwanted observation.  Furthermore, using the Tor router is not a complex endeavor. A user simply connects the router to the Internet via Ethernet or Wi-Fi, then connects his or her device to the router using Wi-Fi.  At which point, the user can now browse the web as they normally would.  On the other hand, setting up a new connection does require a more advanced technical understanding of the router code. However, this can be easily resolved by creating a support program which would handle all of the heavy lifting and optimize configurations for the user.  We also recommend using the Tor Internet browser in tandem with the router, due to the fact that it helps force the user into incorporating safe browsing practices as well as further limiting potential data leaks. However, this is not available to all devices one may wish to secure. 

The primary objective of this work was to explore a privacy solution for the everyday consumer that was cost feasible and effective. This was accomplished by building a Tor router out of a single board computer using open source Tor code. The router successfully hid the user's information from parties outside the barrier of the router, including the ISP. This was essential to the project, as Internet privacy has become a forefront concern. There is always a comprise between efficiency and security; the goal of this work was to provide a solution that is both convenient and effective. The Tor router is portable and can connect to the Internet through both Wi-Fi and Ethernet. While this router did hide a user's information, it came at the price of reduced connection speed. Based on our testing and speed comparisons, users can expect connection speeds that cap at around the 2 Mb/s mark. However, the goal of this project was primarily to aid in the securing of an average user's data - this goal was achieved. Using a Tor router is not a panacea for on-line security, but rather a first line of defense;  We encourage readers to look for additional ways to secure their browsing data.

% or


\appendices


\ifCLASSOPTIONcompsoc
  \section*{Acknowledgments}
\else
  \section*{Acknowledgment}
\fi

We would like to thank Kay Seidel for her original contributions to the group efforts and development of the idea. We would also like to thank the Cybersecurity Education And Research Center for their support and help in the development of this project and paper, as well as urging us to make the leap to a conference rather than letting it be forgotten.

\ifCLASSOPTIONcaptionsoff
  \newpage
\fi



%\bibliographystyle{IEEEtran}

\bibliographystyle{IEEEtran}
\bibliography{tor.bib}

\begin{thebibliography}{10}
\providecommand{\url}[1]{#1}
\csname url@samestyle\endcsname
\providecommand{\newblock}{\relax}
\providecommand{\bibinfo}[2]{#2}
\providecommand{\BIBentrySTDinterwordspacing}{\spaceskip=0pt\relax}
\providecommand{\BIBentryALTinterwordstretchfactor}{4}
\providecommand{\BIBentryALTinterwordspacing}{\spaceskip=\fontdimen2\font plus
\BIBentryALTinterwordstretchfactor\fontdimen3\font minus
  \fontdimen4\font\relax}
\providecommand{\BIBforeignlanguage}[2]{{%
\expandafter\ifx\csname l@#1\endcsname\relax
\typeout{** WARNING: IEEEtran.bst: No hyphenation pattern has been}%
\typeout{** loaded for the language `#1'. Using the pattern for}%
\typeout{** the default language instead.}%
\else
\language=\csname l@#1\endcsname
\fi
#2}}
\providecommand{\BIBdecl}{\relax}
\BIBdecl

\bibitem{snader2008tune}
R.~Snader and N.~Borisov, ``A tune-up for tor: Improving security and
  performance in the tor network.''

\bibitem{reardon2009improving}
J.~Reardon and I.~Goldberg, ``Improving tor using a tcp-over-dtls tunnel,'' in
  \emph{Proceedings of the 18th conference on USENIX security symposium}.\hskip
  1em plus 0.5em minus 0.4em\relax USENIX Association, 2009, pp. 119--134.

\bibitem{rochet2017waterfilling}
F.~Rochet and O.~Pereira, ``Waterfilling: Balancing the tor network with
  maximum diversity,'' \emph{Proceedings on Privacy Enhancing Technologies},
  vol. 2017, no.~2, pp. 4--22, 2017.

\bibitem{kale2016improving}
T.~G. Kale, S.~Ohzahata, C.~Wu, and T.~Kato, ``Improving the tor traffic
  distribution with circuit switching method,'' in \emph{High Performance
  Switching and Routing (HPSR), 2016 IEEE 17th International Conference
  on}.\hskip 1em plus 0.5em minus 0.4em\relax IEEE, 2016, pp. 106--107.

\bibitem{oda2017user}
T.~Oda, M.~Cuka, R.~Obukata, M.~Ikeda, and L.~Barolli, ``A user prediction and
  identification system for tor networks using arima model,'' in
  \emph{International Conference on Emerging Internetworking, Data \& Web
  Technologies}.\hskip 1em plus 0.5em minus 0.4em\relax Springer, 2017, pp.
  89--97.

\bibitem{Konr10cm}
\BIBentryALTinterwordspacing
A.~Forte, N.~Andalibi, and R.~Greenstadt, ``Privacy, anonymity, and perceived
  risk in open collaboration: A study of tor users and wikipedians,'' in
  \emph{Proceedings of the 2017 ACM Conference on Computer Supported
  Cooperative Work and Social Computing}, ser. CSCW '17.\hskip 1em plus 0.5em
  minus 0.4em\relax New York, NY, USA: ACM, 2017, pp. 1800--1811. [Online].
  Available: \url{http://doi.acm.org/10.1145/2998181.2998273}
\BIBentrySTDinterwordspacing

\bibitem{OnionRouting}
M.~G. Reed, P.~F. Syverson, and D.~M. Goldschlag, ``Anonymous connections and
  onion routing,'' \emph{IEEE Journal on Selected Areas in Communications},
  vol.~16, no.~4, pp. 482--494, May 1998.

\bibitem{Konr08cm}
\BIBentryALTinterwordspacing
J.~Wright, ``How tor works,'' 7 1993. [Online]. Available:
  \url{https://jordan-wright.com/blog/2015/02/28/how-tor-works-part-one/}
\BIBentrySTDinterwordspacing

\bibitem{Konr07cm}
\BIBentryALTinterwordspacing
I.~The Tor~Project, ``Tor network: Overview,'' Online. [Online]. Available:
  \url{https://www.torproject.org/about/overview.html.en}
\BIBentrySTDinterwordspacing

\bibitem{Konr06cm}
H.~Neal, ``Onion diagram,'' Creative Commons Attribution-Share Alike 3.0
  Unported License, 3 2008.

\bibitem{alsabah2016performance}
M.~AlSabah and I.~Goldberg, ``Performance and security improvements for tor: A
  survey,'' \emph{ACM Computing Surveys (CSUR)}, vol.~49, no.~2, p.~32, 2016.

\bibitem{Konr09cm}
\BIBentryALTinterwordspacing
I.~The Tor~Project, ``Tips on staying anonymous,'' Online. [Online]. Available:
  \url{https://www.torproject.org/download/download.html.en}
\BIBentrySTDinterwordspacing

\end{thebibliography}

\vfill
\newpage

\section*{About the Authors}

\vspace{-80 mm}
\begin{IEEEbiographynophoto}{Adrian Barberis}
Adrian is a fourth year student at the University of Wyoming.  He is a Computer Science Major and Japanese Language Minor with a love for Machine Learning, Software/Application Development, and Coding in general. He has worked on several projects both in a group and solo setting and is always on the lookout for the next challenge.  Adrian will be graduating from the University of Wyoming in the Spring of 2019.
\end{IEEEbiographynophoto}

\vspace{-80 mm}
\begin{IEEEbiographynophoto}{Dr. Mike Borowczak}
Dr. Mike Borowczak is the Director of the Cybersecurity Education and Research center (CEDAR) and a faculty member of the Computer Science department at the University of Wyoming. He earned his Ph.D. in Computer Science and Engineering (2013) as well as his BS in Computer Engineering (2007) from the University of Cincinnati. His research focuses on detection and prevention of information leakage from hardware side channels. 
\end{IEEEbiographynophoto}

\vspace{-80 mm}
\begin{IEEEbiographynophoto}{Wyatt Emery}
Wyatt is a fourth year student at the University of Wyoming.  He is a Computer Science Major and Japanese Language Minor who will hire on with the Department of Defense, as a programmer, immediately after graduation. He has a passion for programming, problem solving and creating new ideas. He will graduate in May 2018.
\end{IEEEbiographynophoto}

\vspace{-80 mm}
\begin{IEEEbiographynophoto}{Danny Radosevich}
Danny is a fourth year student at the University of Wyoming. He is majoring in Computer Science, and will be graduating in December of 2018. He enjoys learning about computer security, and experiencing new technology.
\end{IEEEbiographynophoto}

\newpage


\enlargethispage{-5in}

\end{document}